\documentclass[twocolumn,showpacs,preprintnumbers,amsmath,amssymb]{revtex4}
\usepackage{pstricks}
\usepackage{color}
\usepackage{epsfig}
\usepackage{graphicx}
\def\beq{\begin{equation}}
\def\be{\begin{eqnarray}}
\def\eeq{\end{equation}}
\def\ee{\end{eqnarray}}

\def\lsim{\buildrel < \over {_{\sim}}}
\def\gsim{\buildrel > \over {_{\sim}}}

\begin{document}

\preprint{RM3-TH/09-6}

\title{Impact of nuclear effects on the determination of the nucleon
axial mass}

\author{Omar Benhar$^1$}
\author{Davide Meloni$^2$}

\affiliation
{
$^1$INFN and Department of Physics\\
``Sapienza'' Universit\`a di Roma, I-00185 Roma, Italy \\
$^2$Department of Physics and INFN\\
Universit\`a ``Roma Tre'', I-00146 Roma, Italy }

\date{\today}
\begin{abstract}
We analize the influence of nuclear effects on the determination of 
the nucleon axial mass from nuclear cross sections.
Our work is based on a formalism widely applied to describe
 electron-nucleus scattering data in the impulse
approximation regime. The results of numerical calculations show that 
 correlation effects, not taken into account by the relativistic 
Fermi gas model, sizably affect the $Q^2$-dependence of the cross section. However, 
their inclusion does not appear to explain the large values of the axial mass
recently reported by the K2K and MiniBooNE collaborations.
\end{abstract}
\pacs{25.30.Pt, 13.15.+g, 24.10.Cn}
\maketitle

Experimental searches of neutrino oscillations exploit neutrino-nucleus interactions 
to detect the beam particles, whose properties are largely unknown.
The use of nuclear targets as detectors,
 while allowing for a substantial increase of the event rate, 
entails non trivial problems, since data analysis requires a quantitative 
understanding of neutrino-nucleus interactions.
In view of the present experimental accuracy, 
the treatment of nuclear effect is in fact regarded as one of the
main sources of systematic uncertainty (see, e.g., Ref.\cite{NUINT07}).

The description of nuclear dynamics is particularly critical to analyses 
aimed at obtaining {\em nucleon} properties from {\em nuclear} cross sections.  

Recently, the K2K \cite{K2K} and MiniBooNE \cite{BOONE} collaborations 
have determined the nucleon axial mass $M_A$, i.e. the mass scale 
driving the $Q^2$-dependence of the dipole parametrization of the nucleon 
axial form factor, from neutrino interactions with oxygen and carbon, 
respectively. The reported value, $M_A \sim 1.2 \ {\rm GeV}$, turn out to be 
significantly larger than the one previously determined from deuterium cross
sections, $M_A \sim 1.0 \ {\rm GeV}$ \cite{oldmass}.
The authors of Ref.\cite{BOONE} argue that the large $M_A$ extracted from the 
data should be regarded as an ``effective axial mass'', embodying nuclear
effects not included in the Relativistic Fermi Gas (RFG) model employed in 
their analysis. They also suggest that replacing the RFG with 
one of the more advanced nuclear models available in the literature 
\cite{benhar1,ahmad,amaro,leitner} may result in 
a value of $M_A$ closer to that measured using deuterium. 

This letter is aimed at assessing the impact of the treatment of nuclear effects 
on the determination of the nucleon axial mass.
Our work is based on the approach described in Refs.\cite{benhar1,benhar2},
in which nucleon-nucleon correlations not included in the RFG model are
 consistently taken into account.

Both K2K and MiniBooNE search for signatures of neutrino oscillations using 
Charged Current Quasi Elastic (CCQE) interactions 
\beq
\nu_\mu + A \rightarrow \mu + p + (A-1) \ , 
\label{process}
\eeq
which are known to yield the dominant contribution to the cross section at neutrino
energy $\lsim 1.5$ GeV \cite{lls}.

The differential cross section of process (\ref{process}), in which a neutrino 
carrying four-momentum $k_\nu=(E_\nu,\bf k_\nu)$ scatters off a nuclear target 
producing a muon of four-momentum $k_\mu=(E_\mu,\bf k_\mu)$, while the target 
final state is undetected, can be written in Born approximation as
\be
\label{xsec}
\frac{d^2\sigma}{d\Omega_\mu dE_\mu}=\frac{G_F^2\,V^2_{ud}}{16\,\pi^2}\,
\frac{|\bf k_\mu|}{|\bf k_\nu|}\,L_{\alpha\beta}\, W_A^{\alpha\beta} \ ,
\ee
where $G_F$ is the Fermi constant and $V_{ud}$ is the CKM matrix element 
coupling $u$ and $d$ quarks. The tensor $L_{\alpha\beta}$ is fully specified by 
the lepton kinematical variables, while the definition of $W_A^{\alpha\beta}$
involves the target initial and final states, as well as the nuclear weak 
current.

For neutrino energies larger than $\sim$ 0.5 GeV, $W_A^{\alpha\beta}$ can be 
obtained within the impulse approximation (IA), i.e. assuming
that neutrino-nucleus scattering reduces to the incoherent sum of scattering 
processes involving individual neutrons, whose momentum (${\bf p}_n$)
and removal energy ($E$) distribution
is described by the nuclear spectral function $P({\bf p}_n,E)$ \cite{spec1,spec2}. 
Neglecting final state 
interactions (FSI) between the struck nucleon and the spectator particles, 
the nuclear tensor can be written in the form \cite{benhar1,benhar2}
\be
W_A^{\alpha\beta}&=& \int d^3p_n\,dE \,P({\bf
p}_n,E)
\  W^{\alpha\beta}_n( {\tilde p}_n,\tilde q) \ ,
\label{hadtensor}
\ee
where $q=k_\nu - k_\mu$ is the four-momentum transfer and
${\tilde p}=({\tilde E}_n,{\bf p}_n)$, 
with ${\tilde E}_n~=~(m_n^2  + |{\bf p}_n|^2)^{1/2}$, $m_n$ 
being the neutron mass. 

The tensor $W^{\alpha\beta}_n$ describes the charged 
current weak interactions of a neutron of initial momentum ${\bf p}_n$
{\em in free space}. The effect of nuclear binding is accounted for by the
replacement $q \rightarrow {\tilde q \equiv (\tilde q_0,{\bf q})}$, with 
\be
\nonumber
\tilde{q}_0  & = & q_0 + M_A - E_{A-1} - {\tilde E}_n  \\ 
&  = & \sqrt{m_p^2 + |{\bf p}_n+{\bf q}|^2}  - \sqrt{m_n^2 + |{\bf p}_n|^2}\ , 
\label{def:qtilde}
\ee
where $M_A$ and $E_{A-1}=[(M_A-m_n+E)^2 + |{\bf p}_n|^2]^{1/2}$ denote 
the target mass and the energy of the recoiling nucleus, respectively \cite{benhar2}. 

The above procedure, originally proposed in the context of a study of electron induced 
nucleon knock out processes \cite{defo}, 
accounts for the fact that a fraction 
of the energy transfer to the target goes into excitation energy of the 
spectator system. The energy $\delta q_0 = q_0 - \tilde q_0$ is spent to put the struck particle 
on the mass shell, and the elementary scattering process is described using free space
kinematics with energy transfer ${\tilde q_0}$.
The physical intepretation of $\tilde q_0$ 
emerges most clearly in the $(|{\bf p}_n|/m_n) \rightarrow 0$ limit, corresponding 
to $\tilde q_0 = q_0 - E$.

The main effect of FSI on the differential inclusive cross section is
a redistribution of the strength, resulting from the coupling of the 
 one particle-one hole final state to more complex n-particle n-hole 
configurations \cite{gangofsix}. This leads to a sizable quenching of the 
cross section in the region of the quasi free peak, corresponding to 
$q_0 \sim Q^2/2m_n$, with $Q^2 = -q^2$, associated with the enhancement of 
the tails at both low and high $q_0$. 

In this work FSI have been described following the approach originally 
developed in Ref.\cite{gangofsix}, in which the inclusive cross section 
is written in the convolution form
\beq
\frac{d\sigma}{d\Omega_\mu dE_\mu} = \int dE_\mu^\prime 
\left( \frac{d\sigma_0}{d\Omega_\mu dE^\prime_\mu} \right)
 f(E_\mu- E^\prime_\mu) \ ,
\end{equation}
where $(d\sigma_0/d\Omega_\mu dE^\prime_\mu)$
is the cross section in the absence of FSI , obtained from Eqs.(\ref{xsec}) and 
(\ref{hadtensor}).
The folding function, embodying
FSI effects, is trivially related to the spectral function of particle states 
\cite{benhar07}. It can be computed using the formalism of 
nuclear many-body theory and the eikonal approximation, i.e.
assuming that: i) the outgoing proton moves along a straight trajectory with
constant speed, and ii) the spectator nucleons act as a collection of fixed 
scattering centers \cite{gangofsix,benhar1}. 

The results of Ref.\cite{benhar1} 
show that inclusion of FSI effects is needed to reproduce the measured 
cross sections of the process $e + ^{16}O \rightarrow e^\prime + X$ at 
electron beam energy $\sim$~1~GeV.

In order to assess the validity of the analysis of Refs.\cite{K2K,BOONE} and the
role of nuclear effects not taken into account by the RFG model, we have
first studied the dependence on $M_A$ of the $Q^2$-distribution of the process
$\nu_\mu + ^{16}O \rightarrow \mu + p + (A-1)$ at beam energy $E_\nu=$1.2~GeV, 
corresponding to the peak in the energy spectrum of the neutrinos used by the 
K2K collaboration \cite{K2K}. 

\begin{figure}[h]
\vspace*{.2in}
\centerline
{\epsfig{figure=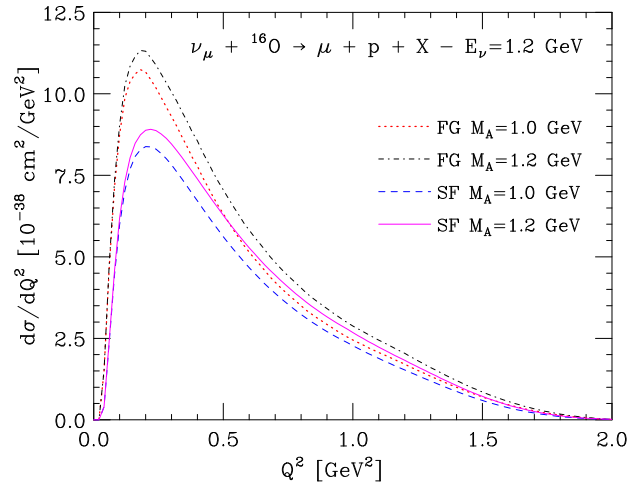,angle=000,width=8.25cm}}
\caption{ (Color online) $Q^2$-dependence of the cross section of the
process $\nu_\mu + ^{16}O \rightarrow \mu + p + X$, for neutrino energy
$E_\nu = 1.2 \ {\rm GeV}$. The dashed and solid lines, labeled SF, have been
obtained using the approach of Refs.\cite{benhar1,benhar2} with $M_A = 1.0$
and $1.2$ GeV, respectively. The corresponding results of the RFG model, 
with Fermi momentum $p_F=$ 225 MeV and removal energy $\epsilon=27$ MeV, are
represented by the dotted and dot-dash lines, labeled FG. \label{fig1} }
\end{figure}

In Fig.~\ref{fig1} we compare the RFG
results to those obtained using Eqs.(\ref{xsec}) and (\ref{hadtensor}) and 
the spectral function of Ref.\cite{spec2}, with $M_A =$ 1.0 and 1.2. 
%
 
It clearly appears that, while increasing the axial mass leads to an enhancement 
of the cross section, the inclusion of correlation effects through the use of a 
realistic spectral function produces a sizable quenching. The dashed line, corresponding 
to the model of Refs.\cite{benhar1,benhar2}, turns out to be below the dot-dash line,
corresponding to the RFG model, over the whole $Q^2$ range. Note that FSI effetcs, 
not taken into account in the calculations, lead to a further suppression of the curves 
labelled SF.

Based on the results of Fig.~\ref{fig1}, we conclude that, as far as the $Q^2$ 
distribution at fixed neutrino energy is concerned, a larger value of the axial 
mass cannot be explained by replacing the RFG with the more advanced model of 
nuclear dynamics discussed in Refs.\cite{benhar1,benhar2}.

It is important to realize, however, that using a realistic momentum and removal
energy distribution may also significanltly affect the determination of $E_\nu$.

From the requirement that the elementary scattering process, 
$\nu_\mu + n \rightarrow \mu + X$, be elastic, i.e. that
\beq
(k_\nu + p_n - k_\mu )^2 = m_p^2 \ ,
\label{onshell:cond}
\eeq
where $m_p$ is the proton mass and the four momentum of the 
struck nucleon is given by $p_n = (E_n,{\bf p}_n)$, with $E_n = M_A-E_{A-1}$, 
it follows that 
\beq  
E_\nu=\frac{m_p^2-m_\mu^2-E_n^2+2E_\mu E_n- 2{\bf k}_\mu \cdot {\bf p}_n+|{\bf p}_n^2|} 
{2( E_n - E_\mu + |{\bf k}_\mu|\cos \theta_\mu - |{\bf p}_n|\cos \theta_n)} \ ,
\label{kin1}
\eeq
where $\theta_\mu$ is the muon angle relative to the neutrino 
beam and $\cos \theta_n = ({\bf k}_\nu \cdot {\bf p}_n)/(|{\bf k}_\nu||{\bf p}_n|)$.

Setting $|{\bf p}_n| = 0$ and fixing the neutron removal energy to 
a constant value $\epsilon$, i.e. setting $E=\epsilon$, implying in turn 
$E_n = m_n - \epsilon$, Eq.(\ref{kin1}) reduces to
\beq
E_\nu = \frac{2E_\mu(m_n - \epsilon)-
(\epsilon^2 - 2 m_n \epsilon + m_\mu^2 + \Delta m^2) }
{2 ( m_n - \epsilon - E_\mu + |{\bf k}_\mu|\cos \theta_\mu )} \ ,
\label{simple:kin}
\eeq
with $\Delta m^2 = m_n^2 - m_p^2$. In the analysis of both K2K and MiniBooNE data, 
the energy of the incoming neutrino has been reconstructed using the above 
equation (compare to Eq.(5) of Ref.\cite{K2K} and Eq.(3) of Ref.\cite{BOONE}), 
 with $\epsilon=$ 27 and 34 MeV for oxygen \cite{K2K} and carbon \cite{BOONE}, respectively. 

Equation(\ref{kin1}) clearly shows that, in general, the knowledge of $E_\mu$ 
and $\theta_\mu$, the 
observables measured by K2K and MiniBooNE, {\em does not} uniquely 
determine the neutrino energy. For any given $E_\mu$ and $\theta_\mu$, $E_\nu$ depends 
on both magnitude and direction of the neutron momentum ${\bf p}_n$, as well as on 
 its removal energy $E$, entering the defintion of $E_n$. 

The distribution of neutrino energy can be obtained from Eq.(\ref{kin1}) using 
values of $|{\bf p}_n|$ and $E$ sampled from the probability distribution 
$|{\bf p}_n|^2 P({\bf p}_n,E)$ and assuming that the polar and azimuthal angles specifying 
the direction of the neutron momentum be uniformly distributed.
\begin{figure}[h]
\vspace*{.2in}
\centerline
{\epsfig{figure=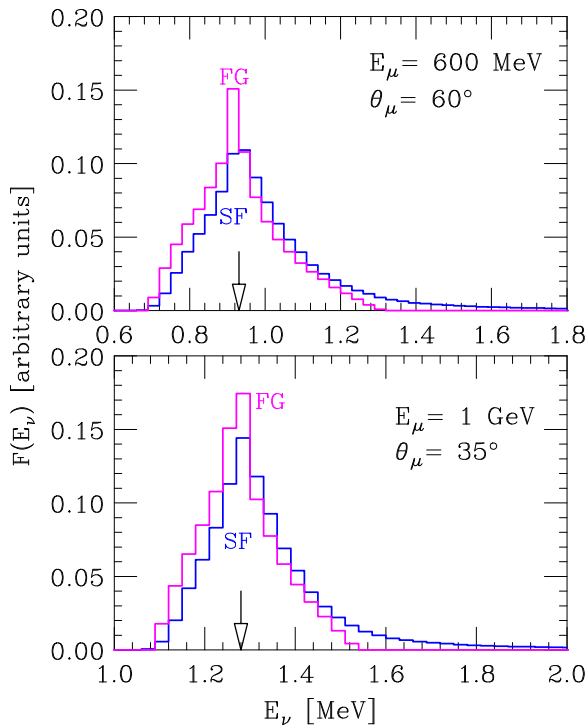,angle=000,width=7.8cm}}
\caption{ (Color online) Upper panel: Neutrino energy distribution at
$E_\mu=$ 600 MeV and $\theta_\mu=$ 60$^\circ$, reconstructed from Eq.(\ref{kin1})
using 2 $\times$10$^4$ pairs of ($|{\bf p}_n|,E$) values sampled from
the probability distributions associated with the oxygen spectral function
of Ref.\cite{spec2} (SF) and the Fermi gas model, with Fermi momentum $p_F=$ 225
MeV and removal energy $\epsilon=27$ MeV(FG). The arrow points to
the value of $E_\nu$ obtained from Eq. (\ref{simple:kin}).
 Lower panel: Same as the upper panel, but
for $E_\mu=$ 1 GeV and $\theta_\mu=$ 35$^\circ$. \label{sampling} }
\end{figure}

The spectral functions of nuclei ranging from carbon to gold, computed in Ref.\cite{spec2},  
exhibit high momentum and high removal energy tails, extending well above 
$|{\bf p}_n|~\gsim$~500~MeV and $ E~\gsim$~100~MeV. 
A direct measurement of the carbon spectral function from the
$(e,e^\prime p)$ cross section at missing momentum and energy up to $\sim$~800
MeV and $\sim~200$ MeV, respectively, has been recently carried out at Jefferson 
Lab \cite{daniela}. The preliminary results of data analysis appear to be consistent 
with the theoretical predictions of Ref.\cite{spec2}.
On the other hand, in the RFG model the typical Fermi momentum, $p_F$, and 
average removal
energy ($\epsilon$ of Eq.(\ref{simple:kin})) are $\sim$~200~MeV and $\sim$~30~MeV, 
respectively.

To gauge the effect of the high momentum and high removal energy tails 
of $P({\bf p}_n,E)$, we have computed the neutrino energy distribution, $F(E_\nu)$, 
using 2 $\times$10$^4$ pairs of ($|{\bf p}_n|,E$) values drawn from the probability 
distributions associated with the oxygen spectral functions of both Ref.\cite{spec2} and 
the RFG model, with Fermi momentum $p_F=$ 225 MeV and removal energy $\epsilon=27$ MeV. 
The results corresponding to $E_\mu=$~600~MeV and $\theta_\mu=$~35$^\circ$, and 
$E_\mu=$~1~GeV and $\theta_\mu=$~35$^\circ$, are displayed in the upper and lower 
panels of Fig.~\ref{sampling}, respectively. 

It appears that the distributions predicted by the RFG model are more sharply
peaked at the neutrino energy given by Eq.(\ref{simple:kin}). On the other hand, the 
$F(E_\nu)$ obtained from the spectral function of Ref.\cite{spec2}
are shifted towards higher energy by $\sim$ 20 MeV, with respect to the RFG results, and 
exhibit a tail extending to very large values of $E_\nu$. 

The average values of $E_\nu$ calculated using the RFG distribution turn out to be 
$\langle E_\nu \rangle =$~1.28~GeV and 934~Mev for the
kinematics of the upper and lower panels, respectively, to be compared with 1.28~GeV and
931~Mev resulting from Eq.(\ref{simple:kin}). On the other hand, the distributions 
associated with the 
spectral functions of Ref.\cite{spec2} yield $\langle E_\nu \rangle =$~1.35~GeV and 
1.01~Gev.

As $Q^2$ grows with $E_\nu$ according to
\beq
Q^2 = 2 E_\nu E_\mu \left(1- \frac{p_\mu}{E_\mu} \cos \theta_\mu \right) - m_\mu^2 \ ,
\label{def:Q2}
\eeq
the results of Fig. \ref{fig1} and \ref{sampling} suggest that using 
neutron energies and momenta obtained from a realistic spectral function in the equation 
determining the kinematics of the CCQE process would lead to extract an even larger value 
of the axial mass.

The energy and momentum distribution of the struck neutron also affects the weak interaction 
vertex, since the neutrino interacts with a {\em bound moving} neutron. As pointed out above, 
within the IA formalism binding is taken into account through a shift of the energy 
transfer. As a result, within the approach of Refs.\cite{benhar1,benhar2} the elementary 
neutrino-neutron scattering process takes place at 
${\tilde Q}^2 = |{\bf q}|^2 - {\tilde q}_0^2 > Q^2$.

In order to assess the full impact of replacing the RFG model with the approach 
of Refs.\cite{benhar1,benhar2}, we have computed the differential cross section
of the process $\nu_\mu + A \rightarrow \mu + p + (A-1)$, as a function of the 
incoming neutrino energy $E_\nu$, for the muon kinematics of Fig. \ref{sampling}.

\begin{figure}[bht]
\vspace*{.2in}
\centerline
{\epsfig{figure=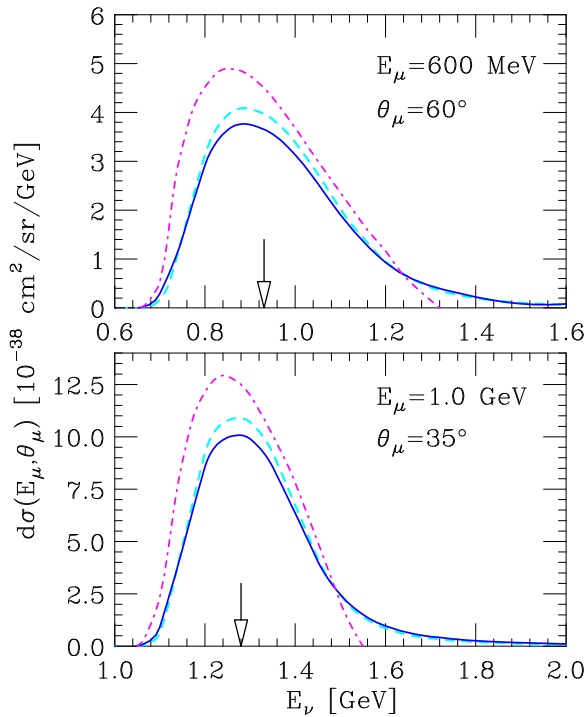,angle=000,width=7.8cm}}
\caption{ (Color online) Upper panel: Differential cross section of the process
$\nu_\mu + A \rightarrow \mu + p + (A-1)$, at $E_\mu=$ 600 MeV
and $\theta_\mu=$ 60$^\circ$, as a function of the incoming neutrino energy.
 The solid line shows the results of the
full calculation, carried out within the approach of Refs. \cite{benhar1,benhar2},
whereas the dashed line has been obtained neglecting the effects
of FSI. The dot-dash line corresponds to the
 RFG model with Fermi momentum $p_F=$ 225 MeV and
removal energy $\epsilon=27$ MeV. The arrow points to
the value of $E_\nu$ obtained from Eq. (\ref{simple:kin}).
Lower panel: Same as the upper panel, but for $E_\mu=$ 1 GeV and
$\theta_\mu=$ 35$^\circ$. \label{nuspec1} }
\end{figure}

The solid lines of Fig.~\ref{nuspec1} show the results of the full calculation, carried out 
using the spectral function of Ref.\cite{spec2}, while the  dashed lines have been obtained 
neglecting the effects of FSI and the dot-dash lines correspond to the RFG model. Note that, 
unlike the histograms of Fig. \ref{sampling}, the curves displayed in Fig. \ref{nuspec1}
have {\rm different} normalizations. As in Fig. \ref{sampling}, the arrows point to the 
values of $E_\nu$ given by Eq.(\ref{simple:kin}).

The differences between the results of the approach of Refs. \cite{benhar1,benhar2} and those
of the RFG model appear to be sizable. The overall shift towards high energies and the 
tails at large $E_\nu$, present in the histograms of Fig. \ref{sampling}, are still clearly 
visible and comparable in size, while the quenching with respect to the RFG model is larger 
than in Fig. \ref{sampling}. Comparison between the dashed and dot-dash lines indicate that 
the $\sim$ 20 \% difference at the peak of the distributions is mainly due to the 
replacement $Q^2 \rightarrow {\tilde Q}^2$, while inclusion of
FSI effects leads to a further reduction of about 8\%.

In conclusion, the results discussed in this paper indicate that nuclear effects not included
in the RFG model significantly affect the $Q^2$-dependence of 
CCQE neutrino-nucleus interactions. However, contrary to the expectation of the authors 
of Ref.\cite{BOONE}, their inclusion does not help to reconcile the large values of 
$M_A$ reported in Refs.\cite{K2K,BOONE} with those extracted from deuterium data.
Using the model of Refs.\cite{benhar1,benhar2} in the data analysis would in fact 
lead to predict an even larger value of the axial mass. 

Other possible explanations of 
the disagreement between the values of $M_A$ obtained by different experiments, such as 
misidentification of CCQE events, should be carefully 
investigated using state-of-the-art models of nuclear structure and dynamics.

While this paper was being drafted, the NOMAD collaboration released the results
of the analysis of quasi-elastic muon neutrino and antineutrino scattering data,
yielding an axial mass $M_A = 1.05 \pm 0.02 (stat) \pm 0.06 (syst)$~GeV \cite{nomad}.
As the NOMAD experiment takes data at beam energies much larger than 
those used in both K2K and MiniBooNE, we have chechek the sensitivity of 
our results to the incoming neutrino energy. It turns out thet the main features of 
the $Q^2$-distribution of the process $\nu_\mu + ^{16}O \rightarrow \mu + p + (A-1)$ 
do not change significantly as $E_\nu$ increases from 1.2 GeV to 10 GeV.

The authors are indebted to M. Sakuda, for drawing their attention to the 
subject of this paper. Useful discussions with L. Ludovici, M.H. Shaevitz and 
M.O. Wascko are also gratefully acknowledged.

\end{document}